# Diagonal free homonuclear 2D correlation using heteronuclei

Bikash Baishya[a]*

[a]*Center of Biomedical Research (Formerly Centre of Biomedical Magnetic Resonance), SGPGIMS Campus, Raebareli Road, Lucknow, 226014, India*

**Abstract**

Homonuclear correlation spectroscopy has played a crucial role in elucidating structure of molecules in solution. In all kinds of homonuclear correlation, the most intense peaks appear in the diagonal, and quite often due to its broad dispersive tails obscure the useful cross peaks close to the diagonal. Herein, we demonstrate that by editing the $^1$H magnetization by a C-13 nucleus in natural abundance before starting the $t_1$ evolution, it is possible to discriminate the inphase $^1$H magnetizations that leads to the diagonal peaks from the antiphase $^1$H magnetizations that generates the cross peaks. The $^1$H magnetization is edited by a C-13 nucleus in natural abundance before starting the $t_1$ period in COSY. After the coherence transfer, the inphase transverse magnetization directly attached to C-13 evolves under one bond heteronuclear coupling for a time period $2\tau$ tuned to one bond heteronuclear coupling while the antiphase transeverse magnetization directly attached to remote C-12 evolves under negligible long range heteronuclear couplings. As a result the inphase magnetization can be eliminated as an unobservable heteronuclear multiple quantum coherence leading to a diagonal free correlated spectrum. The method is demonstrated for COSY.

**Key words:** diagonal free COSY; INEPT, Heteronuclear Multiple Quantum (MQ);

**Introduction:**

Intense diagonal peaks in typical 2D homonuclear correlation spectra quite often obscure the cross peaks near the diagonal. Cross peaks contain the valuable through bond or through space connectivity information between $^1$H spins in a molecule. The intense diagonal peaks obscure nearby cross peaks, lead to strong $t_1$ noise along the indirect dimension, and also reduce the dynamic range of the spectrometer resulting in a lower sensitivity of low intensity cross peaks. Higher the intensity of the diagonal, tougher is the detection of cross peaks close to the diagonal.

A number of methods have been invented for suppressing diagonal peaks. The 1$^{st}$ such approach achieves diagonal peak suppression by recording two spectra-one conventional 2D spectrum, and another containing only the diagonal peaks obtained by setting the mixing time to zero [1-3], and subsequently subtracting the former from the later. This procedure performs well provided relaxation losses are not significant during the mixing time. In addition, accurate reproducibility of the two consecutive experiments must be ensured for efficient suppression of the diagonal, otherwise residual diagonal peaks will appear [2, 4-5].

*\* Corresponding author Fax: 91-522-2668215, E-mail address: bikashbaishya@gmail.com*

In another method diagonal peak suppression is performed by dephasing the magnetization of the excited nucleus by a sequence comprising of defocus, mixing, and refocus [6]. In this method, mixing period is sandwiched between two 90° pulses which are further sandwiched between two separate half of each $t_1$ increment. As a result, the magnetization that does not cross relax during mixing time is rotated by 180°. Subsequently this magnetization is returned to Z, rendering it invisible. The resulting 2D spectrum has unusual appearance as the cross peaks appear as a diagonal with a slope of $\frac{\omega_1}{\omega_2}=2$.

In another elegant approach diagonal suppression is executed by a combination of two jump-and-return sequences before and after the mixing period [7] in 2D NOESY experiment. A pulsed field gradient achieves suppression of magnetization that does not undergo chemical shift change during mixing time.

A novel method of diagonal peak suppression in homonuclear correlated NMR spectra has appeared recently that employs spatial and frequency selective pulses to realize diagonal peak suppression [8]. The method utilizes frequency selective pulses in the presence of weak field gradient to convert a homonuclear system into a spatially separated heteronuclear system [9-15].

In the present work, a method is demonstrated that discriminates the inphase and antiphase magnetization after the mixing period utilizing one bond heteronuclear $^{13}$C-$^{1}$H $J$-coupling in natural abundant samples. The method is applicable to all multidimensional homonuclear correlation experiments.

**Description of the pulse sequences:**

**(a) Diagonal free COSY:** The pulse sequence is displayed in Fig 1A. $G_1$ and $G_2$ gradients are used to destroy any unwanted magnetization from previous scan and to destroy C-13 equilibrium magnetization. After INEPT transfer period from time point 'c' to 'd' of duration $2\tau=1/2[^1J_{CH}]$, where $^1J_{CH}$ is the one bond $^1$H-$^{13}$C heteronuclear scalar coupling, we get the $I_{1Z}S_Z$ spin order at time point 'e' where $I_1$ is proton 1 directly attached to C-13 nucleus 'S' in natural abundance. The transverse magnetization of a neighbouring C-12 attached proton which evolves under very small long range heteronuclear coupling $^2J_{CH}$ or $^3J_{CH}$ is destroyed at this time point by gradient $G_3$. Thus, only C-13 bound protons are selected for the final COSY spectrum. The 90° C-13 pulse after time point 'e' generates C-13 coherence antiphase w.r.to proton $I_1$ i.e. $I_{1Z}S_Y$ at time point 'f'. The 180° pulse on C-13 at time point 'g' flanked by gradients $G_4$ and $G_5$ further ensures removal of any unwanted transverse proton magnetization other than C-13 attached proton. The two simultaneous 90° pulses between time points 'h' and 'i' creates $^1$H coherence antiphase w.r. to C-13 i.e. $I_{1X}S_Z$. Subsequently, from time point 'i' to 'j' tuned to $2\tau=1/2[^1J_{CH}]$], the magnetization is refocused to inphase proton magnetization $I_{1Y}$ w.r.to C-13. Thus the $^1$H magnetization is edited with C-13 before starting the regular COSY $t_1$ evolution period at time point 'j' with $I_{1Y}$ as the starting operator. The C-13 refocusing pulse in the middle of $t_1$ period between 'j' and 'k' ensures refocusing of $^nJ_{C-H}$ coupling where n=1..n bonds. The proton pulse at time point 'o' is the

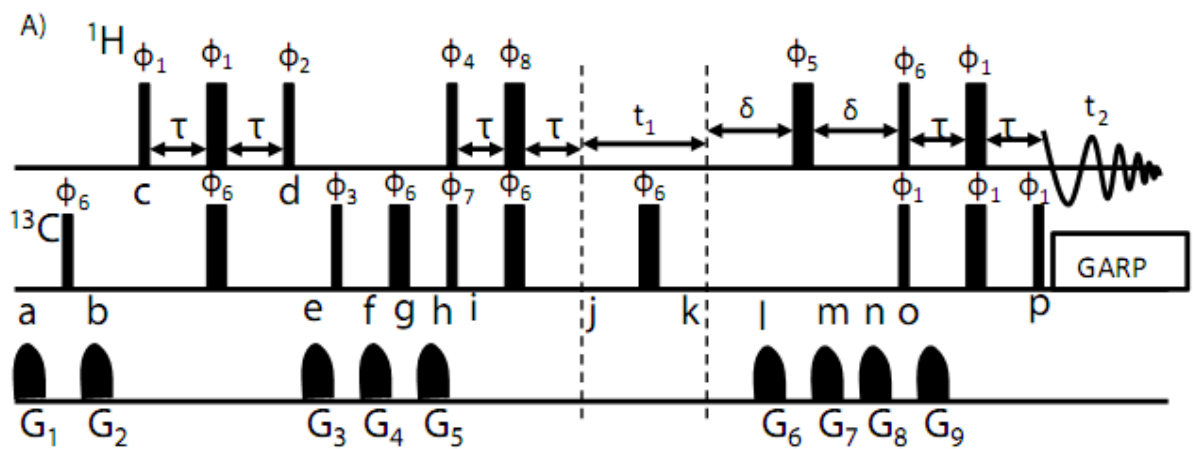

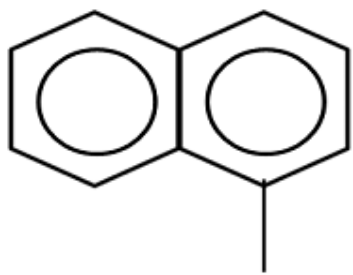

**Figure 1** (A) Diagonal free 2D COSY pulse sequence. The phases are $\Phi_1=0$, $\Phi_2=1$, $\Phi_3=02$, $\Phi_4=1\ 1\ 1\ 1\ 3\ 3\ 3\ 3$, $\Phi_5=0$, $\Phi_6=0$, $\Phi_7=0\ 0\ 2\ 2$, $\Phi_8=0$ and $\Phi_R=02202002$. Thin and fat bars are 90° and 180° pulses. From time point 'c' to 'd' is the INEPT transfer delay. The 90° pulse after time point 'e' creates C-13 coherence i.e. $I_{1Z}S_Y$ at time point 'f'. The 180° pulse on C-13 at time point 'g' flanked by gradients $G_4$ and $G_5$ further selects $^1H$ magnetization attached to C-13 only. The two simultaneous 90° pulses between time points 'h' and 'i' creates $^1H$ coherence $I_{1X}S_Z$. Subsequently, from time point 'i' to 'j' the magnetization is refocused to inphase magnetization $I_{1Y}$ w.r.to C-13. COSY $t_1$ evolution starts at time point 'j' with $I_{1Y}$ as the starting operator. The C-13 refocusing pulse between time points 'j' and 'k' refocus $^nJ_{C-H}$ coupling. The $^1H$ pulse at time point 'o' is the mixing pulse where as the 90° C-13 pulse at time point 'o' converts any unwanted $I_{1X}S_Z$ coherence to unobservable multiple quantum. Gradients $G_8$ and $G_9$ are coherence selection gradients. The 180° pulse between time point 'l' and 'm' refocuses the extra chemical shift evolution due to coherence selection gradient $G_8$ while gradients $G_6$ and $G_7$ in 1:1 ratio removes pulse imperfections associated with this 180° proton pulse. After time point 'o', the antiphase magnetization is $2I_{1Z}I_{2Y}$ and inphase magnetization is $I_{1X}$. Now, $I_{1X}$ evolves under $^1J_{CH}$ from time point 'o' to 'p' and get converted to antiphase coherence $I_{1X}S_Z$ w.r.to C-13, and subsequently removed as an unobservable heteronuclear multiple quantum at time point 'p' by the C-13 90° pulse. $I_{1Z}I_{2Y}$ is the $^1H$ transverse magnetization attached to neighbouring C-12 and its evolution under $^2J_{CH}$ is negligible from time point 'o' 'to 'p'. Therefore, $I_{1Z}I_{2Y}$ leads to cross peak as desired in a correlation experiment. The gradient ratios are

$G_1:G_2:G_3:G_4:G_5:G_6:G_7:G_8:G_9=77:67:80:51:51:53:53:15:15$. (B) Molecular structure of Methyl-1-Naphthalene.

mixing pulse where as the 90° C-13 pulse at time point 'o' converts any unwanted $I_{1X}S_Z$ coherence to unobservable multiple quantum coherence. Due to variation in $^1J_{CH}$ couplings that evolve from 'i' to 'j', small amount of unwanted $I_{1X}S_Z$ coherence may survive during $t_1$ period. The operators present just before the mixing $^1H$ 90° pulse at time point 'o' are

$$-I_{1Y}\cos(\Omega_1 t_1)\cos(\pi J_H t_1) + I_{1X}\sin(\Omega_1 t_1)\cos(\pi J_H t_1) -$$
$$2I_{1X}I_{2Z}\cos(\Omega_1 t_1)\sin(\pi J_H t_1) + 2I_{1Y}I_{2Z}\sin(\Omega_1 t_1)\sin(\pi J_H t_1) + \quad \text{------(1)}$$

Where $I_{2Z}$ is the Z-spin order of proton attached to a neighbouring C-12. After the mixing 90°$_X$ pulse at time point 'o' we get the following terms

$$-I_{1Z}\cos(\Omega_1 t_1)\cos(\pi J_H t_1) + I_{1X}\sin(\Omega_1 t_1)\cos(\pi J_H t_1) +$$
$$2I_{1X}I_{2Y}\cos(\Omega_1 t_1)\sin(\pi J_H t_1) + 2I_{1Z}I_{2Y}\sin(\Omega_1 t_1)\sin(\pi J_H t_1) \quad \text{----------(2)}$$

The 1st and 3rd terms in eq. 2 are unobservable. The second term leads to diagonal peak where as the last term is the desired cross peak. Gradients $G_8$ and $G_9$ are coherence selection gradients for COSY. The 180° pulse between time point 'l' and 'm' refocuses the extra chemical shift evolution due to coherence selection gradient $G_8$ while gradients $G_6$ and $G_7$ in 1:1 ratio removes pulse imperfections associated with this 180° proton pulse between time point 'l' and 'm'.

Diagonal peak suppression is executed after the mixing pulse at time point 'o'. Under normal consequence, the second term in equation 2 will lead to diagonal peak, however, in this case being directly attached to a C-13 nucleus, evolution under one bond heteronuclear coupling for $2\tau=1/2[^1J_{CH}]$ from time point 'o' to 'p' converts it to an antiphase coherence $I_{1Y}S_Z$ w.r.to C-13. On the contrary, $I_{1Z}I_{2Y}$ which normally leads to cross peak, is the proton transverse magnetization attached to neighboring C-12 and its evolution under $^2J_{CH}$ to the C-13 nucleus is negligible for $2\tau=1/2[^1J_{CH}]$, and therefore, remains same $I_{1Z}I_{2Y}$.

Therefore, the terms present just before the 90° C-13 pulse at time point 'p' are

$$I_{1Y}S_Z\sin(\Omega_1 t_1)\cos(\pi J_H t_1) + 2I_{1Z}I_{2Y}\sin(\Omega_1 t_1)\sin(\pi J_H t_1) \quad \text{----------------(3)}$$

The C-13 90° pulse at time point 'p' will convert the 1st term in eq. (3) i.e. $I_{1Y}S_Z$ coherence to an unobservable heteronuclear multiple quantum coherence $I_{1Y}S_Y$. Consequently, only the 2nd term $2I_{1Z}I_{2Y}$ survives resulting in a spectrum containing only the cross peaks. Inspection of the trigonometric coefficient reveals that the cross peaks will have antiphase absorptive lineshape in both dimensions. In overall, the inphase magnetization $I_{1X}$ is converted to unobservable heteronuclear multiple quantum whereas the antiphase magnetization $2I_{1Z}I_{2Y}$ is retained to lead to the desired cross peaks. Signal acquisition occurs under heteronuclear decoupling.

The cross peaks between diastereotopic geminal protons cannot be observed in this diagonal free COSY-because both inphase and antiphase magnetization in eq. 2 are directly attached to the same C-13 nucleus. As a result eq. 3 will have the following form

$$I_{1Y} S_Z \sin(\Omega_1 t_1) \cos(\pi J_H t_1) + 2I_{1Z}I_{2X} S_Z \sin(\Omega_1 t_1) \sin(\pi J_H t_1) \ldots\ldots\ldots\ldots(4)$$

In such a situation, both cross peak and diagonal peak gets converted into unobservable multiple quantum terms after time point 'p'. Therefore, geminal proton-proton correlations for diastereotopic protons do not arise. However, any other neighboring proton to geminal proton correlations appears provided scalar couplings exist between them.

The variations in $^1J_{CH}$ couplings in a real molecule may lead to a weak diagonal in above COSY sequence. The second term $I_{1X}$ in equation 2 does not entirely convert to $I_{1Y}S_Z$ if there exist variations in $^1J_{CH}$ among different spin pairs. Small amount of in-phase magnetization $I_{1X}$ will survive and lead to the presence of a weak diagonal in the diagonal free COSY spectrum.

Similar calculations can be performed for TOCSY and NOESY experiments.

**Experimentals:**

We performed diagonal free 2D COSY experiments on Methyl-1-Naphthalene shown in Figure 1B. Further, regular COSY is also performed under similar conditions for comparison. Experiments were carried out on an 800 MHz NMR spectrometer equipped with a cryoprobe at 300 kelvin. Details of all experimental and processing parameters are given below.

Generally, in poly aromatic compounds COSY cross peak close to the diagonal frequently occurs also in aliphatic side chains in peptides and proteins. Methyl-1-Naphthalene is such a system that has seven aromatic $^1H$ resonances and one methyl resonance. The seven aromatic $^1Hs$ constitutes two independent spin systems of four and three spins respectively, with no noticeable coupling between the two systems. Carrier frequency was put in the middle of aromatic region and all the spectra displayed are for the aromatic region in this molecule.

The diagonal free 2D COSY on Methyl-1-Naphthalene is shown in Figure 2B. It was recorded with $t_1$ and $t_2$ acquisition times of 90 and 250 ms respectively, with States frequency discrimination in $F_1$. Spectral widths were 700 Hz in both dimensions. 2 scan per $t_1$ increment, 8 dummy scans. Relaxation delay was 1.8 sec. Zero filling 1 k and 256 points in $F_2$ and $F_1$ dimensions respectively. Two different processing were done- without any window function, and with unshifted sine bell window function. All these cases are described in results and discussions. The spectrum is displayed after phase correction with a spectral resolution of 2.01 Hz ($F_2$) and 5.55 Hz ($F_1$). The 2τ delays in the pulse sequence of Fig 1B were tuned to 160 Hz, typical of aromatic proton-carbon scalar couplings. Total experimental time was nine minutes

Conventional 2D COSY on Methyl-1-Naphthalene shown in Figure 2A was recorded with same parameters as diagonal free COSY above with echo antiecho frequency discrimination in $F_1$.

**Results and discussions:**

**(a) Diagonal free COSY:**

Figure 2A and 2B displays the comparison of the regular 2D COSY and diagonal free 2D COSY respectively recorded on Methyl-1-Naphthalene under identical conditions. The diagonal peaks are completely suppressed in 2B. Both experiments were recorded with same number of scans, $F_1$ and $F_2$ resolutions, and processed without any resolution enhancing window function to reduce sensitivity loss. Details of experimental and processing parameters are given in experimental section. In regular COSY of Figure 2A, the two cross peaks (marked as 3 and 4) close to the diagonal peaks (marked as 1 and 2) are obscured shown expanded in Fig. 2C. Assignment of cross peak is tedious in such a situation. However, the diagonal free COSY spectrum in Figure 2B clearly displays the two cross peaks 3 and 4 due to the absence of diagonal peaks 1 and 2. Peaks 3 and 4 from 2B are shown expanded in Figure 2D with absorptive antiphase line shape in both dimensions.

Strongly resolution enhancing unshifted sine bell window function is utilized in conventional 2D COSY to improve cross peaks close to the diagonal, and the procedure performs well for peaks which are not too close to the diagonal. As a drawback, loss of sensitivity is observed. Figure 2E and 2F shows same cross peaks of 2C and 2D processed with unshifted sine bell function in both dimensions. Cross peaks 3 and 4 in 2E, which are very close to the diagonal peaks 1 and 2 are still obscured although the situation is better than 2C. Obviously the diagonal free version in 2F shows great clarity in cross peak identification as the diagonal peaks 1 and 2 are absent. Thus, the diagonal free COSY experiments detects cross peak very close to the diagonal without the need of any resolution enhancing unshifted sine bell function.

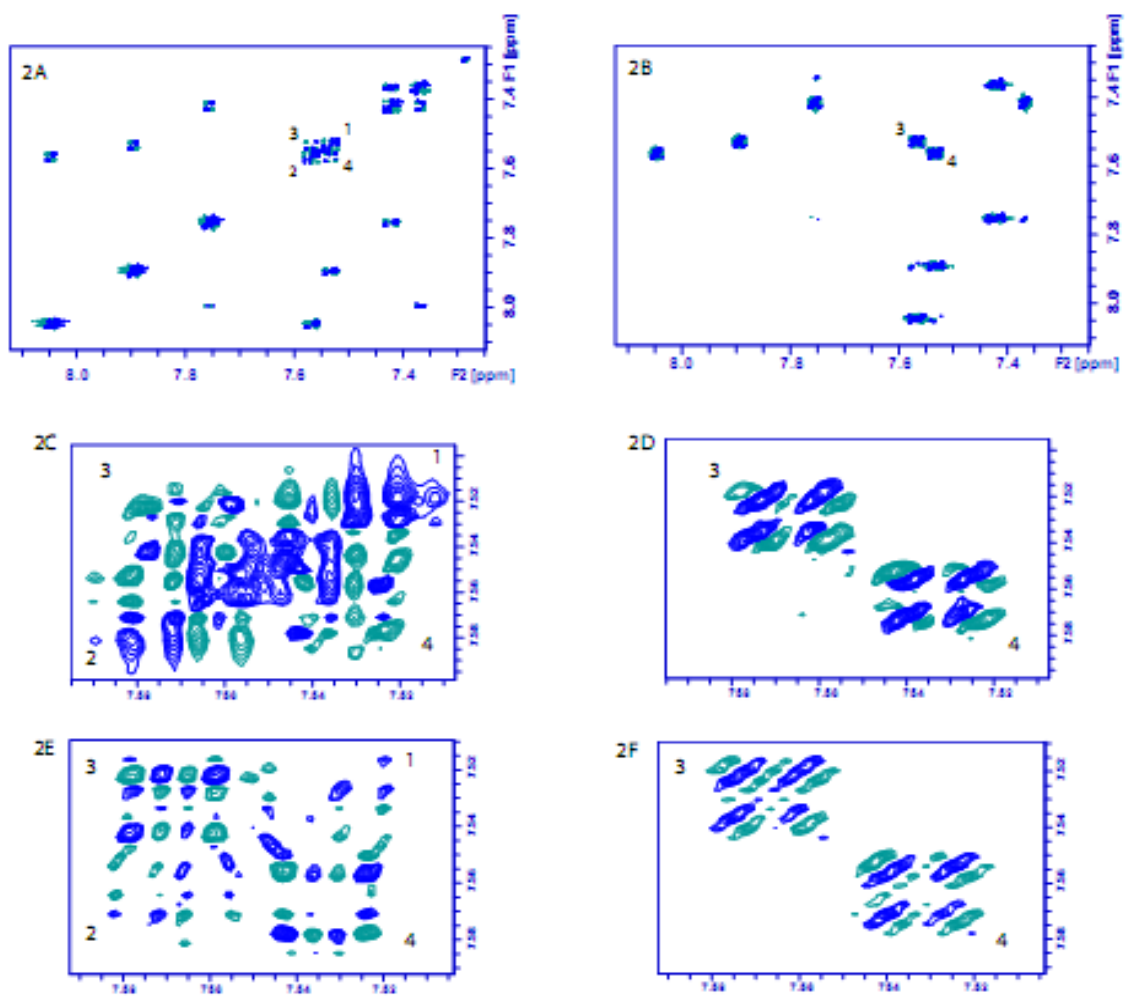

**Figure 2.** Regular 2D COSY and diagonal free 2D COSY are displayed in A and B respectively on Methyl-1-Naphthalene for same acquisition and processing parameters without any window function. Absence of diagonal peaks is clearly visible in 2B. In Figure 2A, the two cross peaks (marked as 3 and 4) close to the diagonal peaks (marked as 1 and 2) are obscured with poor visibility shown expanded in Fig. 2C. However, the diagonal free COSY spectrum in Fig 2B clearly displays the two cross peaks 3 and 4 shown expanded in Fig 2D with absorptive antiphase line shape in both dimensions. Figure 2E and 2F shows same regions of 2C and 2D processed with strongly resolution enhancing unshifted sine bell function in both dimensions. Cross peaks 3 and 4 in 2E, which are very close to the diagonal peaks 1 and 2 are still not properly visible although the situation is better than that in Figure 2C due to the application of unshifted sine bell window function. Obviously the diagonal free version in 2F shows great clarity in cross peak identification as the diagonal peaks 1 and 2 are absent.

**Conclusion:**

Diagonal free correlation spectroscopy of protons utilizing natural abundant C-13 nuclei is demonstrated. The method is in principle applicable to any homonuclear correlation experiments. Significant sensitivity loss occurs due to filtration of the $^1$H signals through natural abundant heteronuclei, however some part can be recovered by increasing receiver gain due to the absence of strong diagonal signals. Although sensitivity of the diagonal free experiment is less, the experiment of Figure 2B on methyl-1-Naphthalene could be recorded in nine minutes at 800 MHz using cryoprobe. Therefore for small molecules these are highly practical. Unlike Glanzer and Zaggner method, the current method does not utilize any long spatially and spectrally selective pulses in the presence of weak gradient to achieve diagonal peak suppression. Further, optimization of such pulses may not be easy. The method presented here simply employs very short period of evolution under one bond C-H coupling which is typically 3 to 4 ms long in most of the molecules, and therefore quite user friendly. Detection of cross peaks close to the diagonal is more visible when $t_1$ acquisition time is small 20 to 30 ms. Thus, a novel method is demonstrated for assignment of cross peaks close to the diagonal peaks. Applications to other experiments are on progress.